\documentclass[a4paper]{JHEP3}

\usepackage{epsfig,multicol,bbm}

\usepackage{amssymb,amsmath}
\usepackage{braket}
\usepackage{graphicx}

\newcommand\DCI{\ensuremath{D_{\textrm{\scriptsize CI}}}}
\newcommand\order[1]{\ensuremath{\mathcal{O}\left(#1\right)}}
\newcommand\unit{\mathbbm{1}}
\newcommand\comm[2]{\ensuremath{\left[#1,#2\right]}}

\newcommand\tr{\ensuremath{Tr}}
\newcommand\E[1]{\ensuremath{\mathrm{e}^{#1}}}
\newcommand\MSbar{\ensuremath{\overline{\textrm{MS}}}}
\newcommand\RI{\ensuremath{\textrm{RI}}}
\newcommand\RIprime{\ensuremath{\textrm{RI'}}}
\newcommand\RGI{\ensuremath{\textrm{RGI}}}
\newcommand\Eqref[1]{Eq.\;(\ref{#1})}

\newcommand\Tabref[1]{Table\;\ref{#1}}
\newcommand\Tabsref[2]{Tables\;\ref{#1}-\ref{#2}}

\newcommand\Figref[1]{Fig.\;\ref{#1}}

\newcommand\Secref[1]{Sect.\;\ref{#1}}
\newcommand\MeV{\ensuremath{\,\textrm{MeV}} }
\newcommand\GeV{\ensuremath{\,\textrm{GeV}} }
\newcommand\fm{\ensuremath{\,\textrm{fm}} }
\newcommand\mAWI{\ensuremath{m_{\textrm{\scriptsize AWI}}}}
\newcommand\const{\ensuremath{\textrm{const.}}}

\newcommand\ie{i.~e.~}
\def\plotwidth{0.9\textwidth}
\def\plotwidthsmall{0.82\textwidth}

\title{Renormalization factors of quark bilinears using the \DCI\, operator with dynamical quarks}

\author{Philipp Huber\\
	Institut f. Physik, FB Theoretische Physik, Universit\"at Graz\\
	E-mail: \email{philipp.huber@uni-graz.at}}

\abstract{Non-perturbative renormalization factors of bilinear quark operators are computed for the Chirally Improved lattice action with two dynamic quarks. The analysis is based on five different parameter sets with lattice size $12^3\times24$ and four parameter sets with lattice size $16^3\times32$. For the pseudoscalar renormalization factor the pion pole contribution is subtracted and chiral extrapolations are performed. Results are given in \RIprime- and \MSbar-scheme as well as in \RGI-form.}


\preprint{}

\begin{document}

\section{Motivation}\label{sec:motivation}

In order to determine renormalized quantities like quark masses, the chiral condensate or the pion decay constant on the lattice one needs renormalization factors to compare with results from experiments or continuum theories. Those values are typically given in the modified minimal subtraction (\MSbar) renormalization scheme.\\
Lattice Dirac operators satisfying the Ginsparg-Wilson (GW) equation \cite{Ginsparg:1982bj} implement a version of chiral symmetry that is closest to the continuum form, with only local $\order{a}$ violations. Exact GW-operators have the benefit of relations that simplify the renormalization procedure \cite{Luscher:1998pq}. For Dirac operators that fulfill the GW-equation only approximately computing renormalization factors also provides a check how well chiral symmetry is realized in this formulation.\\
Let us consider local, flavor non-singlet bilinear quark-field operators
\begin{equation}\label{eq:motivation_local_operator}
  O_\Gamma = \overline{u} \, \Gamma \, d \, ,
\end{equation}
where $\Gamma$ denotes a matrix from the Clifford algebra in the chiral representation and $u$ and $d$ denote the quark fields for the light quarks. We organize the 16 matrices into the scalar, vector, tensor, axial vector and pseudoscalar sectors with $S \sim \unit, V \sim \gamma_\mu, T \sim \frac{i}{2} \comm{\gamma_\mu}{\gamma_\nu}, A \sim \gamma_\mu \gamma_5 , P \sim \gamma_5$ according to their properties under Lorenz transformations. For a chiral Dirac operator we have $Z_P = Z_S$ and $Z_A = Z_V$. For conserved covariant currents we have $Z_A = 1$ due to Ward identities. \\
Although only the overlap operator \cite{Narayanan:1993ss,Narayanan:1992wx} satisfies the GW-equation exactly, there is a number of formulations that have good chiral properties by satisfying the GW-constraint approximately (fixed point operator \cite{Hasenfratz:1993sp}, \DCI\,\cite{Gattringer:2000js}) or in some limit (domain wall fermions \cite{Kaplan:1992bt,Furman:1994ky}). Within the BGR-collaboration fixed point and CI operators have been studied in both, the quenched approximation \cite{Gattringer:2003qx} and the CI operator was analyzed in full QCD in \cite{Lang:2005jz,Lang:2005im,Gattringer:2008vj}. \\
The \DCI\, has been introduced in \cite{Gattringer:2000js,Gattringer:2000qu} as an approximate solution to the GW-equation, where the lattice derivative operator does not only include nearest neighbor interactions, but also more remote connection, each of which carries a Dirac structure.\\
Renormalization factors for the quenched case have been computed in \cite{Gattringer:2004iv} and this work follows the procedure therein closely. Non-perturbative renormalization in the quenched approximation has also been performed for the Wilson operator \cite{Gimenez:1998ue,Donini:1999sf,Giusti:1998gx,Becirevic:1998yg}, staggered fermions \cite{Ishizuka:1998ji}, domain-wall fermions \cite{Blum:2001sr} and the overlap operator \cite{Giusti:2001pk,Zhang:2005sca,DeGrand:2005af}. Calculations with dynamic fermions were performed non-pertubatively for the overlap operator in \cite{Noaki:2009xi}. For dynamical clover fermions perturbative results were presented in \cite{Skouroupathis:2008mf,Skouroupathis:2010mq}, while \cite{Gockeler:2010yr} covers perturbative and non-perturbative approaches.\\
In \Secref{sec:method} the method to determine renormalization constants \cite{Martinelli:1995ty} is reviewed. Technicalities like the parameters used for generating gauge configurations and data analysis are dealt with in \Secref{sec:technicalities}. Results in different schemes are discussed in \Secref{sec:results_ri_prime}, \Secref{sec:results_ms} and \Secref{sec:results_rgi} and we summarize our results and conclude in \Secref{sec:summary_and_conclusion}.


\section{Method}\label{sec:method}

We want to compute non-perturbative renormalization constants on the lattice using a scheme applicable in both, lattice simulations and perturbative continuum calculations. To this end we use the regularization independent scheme (\RI) proposed in \cite{Martinelli:1995ty}. Results in the \RI-scheme can also be converted to the more conventional \MSbar-scheme in a perturbative way.
In the \RI-scheme we match expectation values of bilinear quark operators between quark fields at a specific momentum value $p^2=\mu^2$ with the renormalization scale $\mu$,
\begin{equation}\label{eq:method_operator_sandwich}
  \left. \Braket{p|O_\Gamma|p} \right|_{p^2=\mu^2}
\end{equation}
to corresponding tree-level matrix elements $\braket{p|O_\Gamma|p}_0$. Continuum perturbation theory can only be applied for a renormalization scale much larger than the QCD scale parameter $\Lambda_{\textrm{QCD}}$. Discretization effects can be neglected for $\mu$ much smaller than the lattice cut-off $1/a$ so the renormalization procedure is expected to work in a window
\begin{equation}\label{eq:method_renormalization_window}
  \Lambda_{\textrm{QCD}}^2 \ll \mu^2 \ll 1/a^2 \, .
\end{equation}
Expression \eqref{eq:method_operator_sandwich} is gauge-variant so we need to fix the gauge in order to compare results obtained in a lattice formulation to continuum results. Landau gauge can be implemented in both approaches, but has the problem of Gribov ambiguities. In \cite{Paciello:1994gs,Giusti:2001xf,Giusti:2002rn,Gattringer:2004iv} this has been addressed and no significant effects were detected, therefore no explicit check was performed in this work.\\
We use the method from \cite{Martinelli:1995ty} with the modifications from \cite{Gockeler:1998ye} to compute renormalization factors.
Multiplying \eqref{eq:method_operator_sandwich} with the inverse of the tree-level matrix element $\braket{p|O_\Gamma|p}_0^{-1}$ gives us the renormalization condition
\begin{equation}\label{eq:method_renormalization_condition}
  Z_\Gamma \frac{1}{12} \tr \left. \left[ \Braket{p|O_\Gamma|p} \Braket{p|O_\Gamma|p}_0^{-1} \right] \right|_{p^2=\mu^2} = 1 \, .
\end{equation}
The matrix element
\begin{equation}\label{eq:method_bilinear_matrix_element}
  \Braket{p|O_\Gamma|p} = \frac{1}{Z_q} \Lambda_\Gamma(p)
\end{equation}
with the quark field renormalization factor $Z_q$ is proportional to the amputated Green function
\begin{equation}\label{eq:method_amputated_green_function}
  \Lambda_\Gamma(p) = S^{-1}(p) G_\Gamma(p) S^{-1}(p) \, .
\end{equation}
The inverse quark propagator is denoted $S^{-1}(p)$. The Green function $G_\Gamma(p)$ is computed by taking the expectation value of the local operator $O_\Gamma(z)$ and Fourier transforming it. This reads
\begin{equation} \label{eq:method_green_function_basic}
  G_\Gamma(p)_{\alpha,\beta} = \frac{1}{V} \sum_{x,y} \E{-i p (x-y)} \Braket{u_\alpha(x)\sum_z O_\Gamma(z) \overline{d}_\beta(y)} \, ,
\end{equation}
where the indices $\alpha, \beta$ run over color and Dirac indices and $V$ denotes the lattice volume. We use equal masses for $u$ and $d$ quarks, hence the quark propagator is
\begin{equation}
  S_{\alpha,\beta}(x, y) = \Braket{u_\alpha(x)\overline{u}_\beta(y)} = \Braket{d_\alpha(x)\overline{d}_\beta(y)} \, .
\end{equation}
Using the modification \cite{Gockeler:1998ye} we compute quark propagators with momentum sources
\begin{equation} \label{eq:method_quark_propagator_momentum_space}
  S_n(x|p) = \sum_y \E{i p y} S_n(x,y)
\end{equation}
for each gauge configuration. Inserting the local operator defined in \Eqref{eq:motivation_local_operator} into \Eqref{eq:method_green_function_basic} we can rewrite $G_\Gamma(p)$ in terms of the quark propagator $S(x|p)$. Using $\gamma_5$-hermiticity, we rewrite the Green function
\begin{equation}
\begin{split}
  G_\Gamma(p) &=       \frac{1}{V} \sum_{x,y,z} \E{-i p (x-y)} \Braket{u_\alpha(x)\overline{u}_\beta(z) \Gamma d_\alpha(z)\overline{d}_\beta(y)} \\
              &\approx \frac{1}{V N} \sum_{n=1}^N \sum_z \gamma_5 S_n(z|p)^\dagger \gamma_5 \Gamma S_n(z|p) \, ,
\end{split}
\end{equation}
where $N$ is the number of gauge configurations we average over. The quark propagator in momentum space is obtained by transforming \Eqref{eq:method_quark_propagator_momentum_space} fully to momentum space
\begin{equation}
  S(p) \approx \frac{1}{V N} \sum_{n=1}^N \sum_x \E{-i p x} S_n(x|p) \, .
\end{equation}
We use momentum sources \cite{Gockeler:1998ye} with momenta listed in \Tabref{tab:parameters_momentum_vectors_12x24} and \Tabref{tab:parameters_momentum_vectors_16x32} in order to compute $S_n(y|p)$ by solving
\begin{equation} \label{eq:method_momentum_source_equation}
  \sum_y \DCI(x,y) S_n(y|p) = \E{i p x}
\end{equation}
for the chirally improved Dirac operator \DCI. This implies we have to solve \Eqref{eq:method_momentum_source_equation} for every momentum vector, but improves the signal significantly. In momentum space this Dirac operator reading
\begin{equation}\label{eq:method_dci_construction}
  \DCI(p) = R(p) + i \gamma_\mu a_\mu(p)
\end{equation}
in the free massless case is constructed using the functions \cite{Gattringer:2000js,Gattringer:2000qu}
\begin{equation}
  a_\mu(p) = i p_\mu + \order{a p}^2 \quad \textrm{and} \quad
  R(p)     = \order{a p}^2 \, .
\end{equation}
The quark field renormalization is obtained in the so-called \RIprime-scheme, where we can apply \Eqref{eq:method_dci_construction} and find
\begin{equation}
\begin{split}
  Z_q^{\RIprime}(p) &= \left. \frac{1}{12} \tr S^{-1}(p) S_0(p) \right|_{p^2=\mu^2} \\
                    &= \left. \frac{1}{12} \tr \left( S^{-1}(p) \frac{R(p) - i\,a_\mu(p) \gamma_\mu}{R(p)^2 + a_\mu(p) a_\mu(p)} \right) \right|_{p^2=\mu^2} \, .
\end{split}
\end{equation}
Renormalization factors for quark bilinears can then be computed from the renormalization condition \Eqref{eq:method_renormalization_condition} together with the definitions \Eqref{eq:method_bilinear_matrix_element} and \Eqref{eq:method_amputated_green_function} for the \RIprime-scheme
\begin{equation}\label{eq:method_Z_Gamma_definition}
\begin{split}
  Z_\Gamma^{\RIprime}(p) &= \left. \frac{12}{\tr \left[ \Braket{p|O_\Gamma|p} \Braket{p|O_\Gamma|p}_0^{-1} \right]} \right|_{p^2=\mu^2} \\
                         &= \left. \frac{12 Z_q^{\RIprime}(p)}{\tr \left[ S^{-1}(p) G_\Gamma(p) S^{-1}(p) \Gamma^{-1} \right]} \right|_{p^2=\mu^2} \, ,
\end{split}
\end{equation}
where we use $\Braket{p|O_\Gamma|p}_0 = \Gamma$.
For the vector, axial vector and tensor sectors we average over the components before taking the trace.\\
A modified version of the method described above was presented in \cite{Capitani:2000xi} and also in \cite{Maillart:2008pv}, which will be referred to as the ``reduced'' method. The contribution that comes with the unit matrix of the quark propagator introduces a pure cut-off effect and should therefore be eliminated from the renormalization prescription. A way to achieve this is to define a reduced quark propagator
\begin{equation} \label{eq:quark_propagator_definition_reduced}
  \overline{S}(p) = S(p) - \left( \frac{1}{12} \tr S(p) \right) \unit \, ,
\end{equation}
as well as a reduced free quark propagator
\begin{equation}
  \overline{S}_0(p) = S_0(p) - \left( \frac{1}{12} \tr S_0(p) \right) \unit \, .
\end{equation}
Consequently a reduced quark field renormalization factor
\begin{equation}
  \overline{Z}_q^{\RIprime}(p) = \left. \frac{1}{12} \tr \overline{S}^{-1}(p) \overline{S}_0(p) \right|_{p^2=\mu^2}
\end{equation}
can be computed. Analogously \Eqref{eq:method_Z_Gamma_definition} can be redefined in these terms
\begin{equation} \label{eq:Z_Gamma_definition_reduced}
  \overline{Z}_\Gamma^{\RIprime}(p) = \left. \frac{12 \overline{Z}_q^{\RIprime}(p)}{\tr \left[ \overline{S}^{-1}(p) \hat{G}_\Gamma(p) \overline{S}^{-1}(p) \Gamma^{-1} \right]} \right|_{p^2=\mu^2} \, ,
\end{equation}
with
\begin{equation}
  \hat{G}_\Gamma(p)\approx \frac{1}{V N} \sum_{n=1}^N \sum_z \gamma_5 \overline{S}_n(z|p)^\dagger \gamma_5 \Gamma \overline{S}_n(z|p) \, .
\end{equation}
We state the final numbers for the renormalization factors in both, the non-reduced and the reduced scheme.



\section{Technicalities}\label{sec:technicalities}



\subsection{Parameters}\label{sec:technicalities_parameters}


Gauge configurations were generated with two dynamic quarks using the L\"uscher-Weisz gauge action \cite{Luscher:1985zq} and stout smearing \cite{Morningstar:2003gk}, which are presented in \cite{Lang:2005jz,Lang:2005im,Gattringer:2008vj}. Prior results in the quenched approximation utilized hypercubic smearing instead of stout smearing, hence a comparison of renormalization factors for the quenched and the dynamic case are not applicable. The gauge couplings and sea-quark masses are listed in \Tabref{tab:parameters_lattice_configs}. For the $12^3\times24$ setups we compute 10, for the $16^3\times32$ setups 5 momentum propagators for each momentum vector.
\TABLE{
\begin{tabular}{c|rcr|cc|cc|c}
run & $ L^3\times T $   & $\beta$ & $a\, m_q$  & $a\, m_{\textrm{\scriptsize AWI}}$ & $m_{\textrm{\scriptsize AWI}} [MeV]$ & $a[\textrm{fm}]$ & $a[\textrm{GeV}^{-1}]$ & \#\,cf. \\ \hline
a   & $12^3\times24$    & $4.7$   & $-0.05$  & $0.023(2)$                       & $30.9(8)$                            & $0.147(18)$      & $0.75(9)$              & 10 \\
b   & $12^3\times24$    & $5.2$   & $ 0.02$  & $0.025(1)$                       & $42.1(2)$                            & $0.115(6)$       & $0.58(3)$              & 10 \\
c   & $12^3\times24$    & $5.2$   & $ 0.03$  & $0.037(1)$                       & $58.1(2)$                            & $0.125(6)$       & $0.63(3)$              & 10 \\
d   & $12^3\times24$    & $5.3$   & $ 0.04$  & $0.037(2)$                       & $60.8(2)$                            & $0.120(4)$       & $0.61(2)$              & 10 \\
e   & $12^3\times24$    & $5.3$   & $ 0.05$  & $0.050(19)$                      & $76.4(2)$                            & $0.129(1)$       & $0.654(5)$             & 10 \\ \hline
f   & $16^3\times32$    & $4.65$  & $-0.06$  & $0.02664(15)$                    & $35.05(19)$                          & $0.150(1)$       & $0.760(5)$             & 5 \\ 
g   & $16^3\times32$    & $4.70$  & $-0.05$  & $0.03277(31)$                    & $43.11(41)$                          & $0.150(2)$       & $0.76(1)$              & 5 \\ 
h   & $16^3\times32$    & $4.58$  & $-0.077$ & $0.01097(26)$                    & $15.03(36)$                          & $0.144(2)$       & $0.73(1)$              & 5 \\ 
i   & $16^3\times32$    & $4.65$  & $-0.07$  & $0.00861(36)$                    & $12.09(52)$                          & $0.140(1)$       & $0.711(5)$             & 5 \\
\end{tabular}
\caption{\label{tab:parameters_lattice_configs}Parameters of the different runs with AWI mass and lattice spacing measurements thereon. The lattice spacing and AWI mass values are taken from \cite{Gattringer:2008vj,Lang:2005jz}. For run $i$ the values are preliminary and not yet published. Runs $a-e$ denote runs on the smaller $12^3\times24$ lattices, while runs $f-i$ denote the runs on the $16^3\times32$ lattices. Note that the lattice spacing $a$ for run $a$ is closer to runs with big lattice size than to the remaining four $12^3\times24$ lattices.}
}
\TABLE{
\begin{tabular}{cccc|l|lllll}
          &           &           &           &             & a            & b            & c            & d            & e          \\
$a\, p_1$ & $a\, p_2$ & $a\, p_3$ & $a\, p_4$ & $a\, p$     & $p [\GeV]$   & $p [\GeV]$   & $p [\GeV]$   & $p [\GeV]$   & $p [\GeV]$ \\ \hline
0 & 0 & 0 & 0 & 0.1309 & 0.1757 & 0.2246 & 0.2066 & 0.2153 & 0.2002 \\
1 & 0 & 0 & 0 & 0.3927 & 0.5271 & 0.6738 & 0.6199 & 0.6458 & 0.6007 \\
0 & 0 & 0 & 1 & 0.5397 & 0.7245 & 0.9261 & 0.8520 & 0.8875 & 0.8256 \\
1 & 0 & 0 & 1 & 0.6545 & 0.8786 & 1.1230 & 1.0332 & 1.0763 & 1.0012 \\
0 & 1 & 1 & 1 & 0.9163 & 1.2300 & 1.5723 & 1.4465 & 1.5068 & 1.4016 \\
1 & 1 & 1 & 1 & 0.9883 & 1.3266 & 1.6958 & 1.5601 & 1.6251 & 1.5117 \\
2 & 1 & 1 & 1 & 1.1184 & 1.5013 & 1.9191 & 1.7655 & 1.8391 & 1.7108 \\
3 & 1 & 1 & 1 & 1.2892 & 1.7306 & 2.2121 & 2.0352 & 2.1200 & 1.9721 \\
2 & 1 & 1 & 2 & 1.4399 & 1.9329 & 2.4707 & 2.2730 & 2.3678 & 2.2026 \\
3 & 1 & 1 & 2 & 1.5762 & 2.1159 & 2.7047 & 2.4883 & 2.5920 & 2.4111 \\
4 & 2 & 2 & 2 & 2.1628 & 2.9033 & 3.7112 & 3.4143 & 3.5565 & 3.3084 \\
\end{tabular}
\caption{\label{tab:parameters_momentum_vectors_12x24}The available momentum combinations on the $12^3 \times 24$ lattice. Parameters of the runs are listed in \Tabref{tab:parameters_lattice_configs}.}
}
\TABLE{
\begin{tabular}{cccc|l|llll}
          &           &           &           &             & f            & g            & h            & i           \\
$a\, p_1$ & $a\, p_2$ & $a\, p_3$ & $a\, p_4$ & $a\, p$     & $p [\GeV]$   & $p [\GeV]$   & $p [\GeV]$   & $p [\GeV]$  \\ \hline
2 & 1 & 1 & 1 & 0.8388 & 1.1011 & 1.1035 & 1.1494 & 1.1802 \\
3 & 1 & 1 & 1 & 0.9669 & 1.2693 & 1.2720 & 1.3250 & 1.3604 \\
2 & 1 & 1 & 2 & 1.0799 &        &        & 1.4798 & \\
4 & 1 & 1 & 1 & 1.1151 &        &        & 1.5280 &        \\
3 & 1 & 1 & 2 & 1.1822 & 1.5519 & 1.5552 & 1.6200 & 1.6633 \\
4 & 1 & 1 & 2 & 1.3061 & 1.7146 &        & 1.7898 &        \\
3 & 1 & 2 & 2 & 1.3639 & 1.7904 & 1.7942 & 1.8690 & 1.9189 \\
3 & 2 & 2 & 2 & 1.5241 & 2.0007 & 2.0049 & 2.0885 & 2.1443 \\
4 & 2 & 2 & 2 & 1.6221 & 2.1294 & 2.1339 & 2.2228 & 2.2822 \\
6 & 1 & 2 & 2 & 1.7369 & 2.2801 & 2.2849 & 2.3801 & 2.4437 \\
5 & 1 & 2 & 3 & 1.8235 & 2.3938 & 2.3989 & 2.4988 & \\
5 & 2 & 2 & 3 & 1.9462 & 2.5549 & 2.5603 & 2.6670 & 2.7383 \\
\end{tabular}
\caption{\label{tab:parameters_momentum_vectors_16x32}The available momentum combinations on the $16^3 \times 32$ lattice. The $p$ values are only listed if we computed the quark propagator for this momentum vector. Parameters of the runs are listed in \Tabref{tab:parameters_lattice_configs}.}
}

\TABLE{
\begin{tabular}{c|r|rrrrrrrrrr}
run & $a\, m_q$   & \multicolumn{5}{c}{$a\, m_q$}  \\ \hline
a   & $-0.05$  & $-0.025$  & $0.0$     & $0.03$    & $0.06$   & $0.10$  \\
b   & $ 0.02$  & $0.065$   & $0.11$    & $0.155$   & $0.20$  \\
c   & $ 0.03$  & $0.0725$  & $0.115$   & $0.1575$  & $0.20$  \\
d   & $ 0.04$  & $0.08$    & $0.12$    & $0.16$    & $0.20$  \\
e   & $ 0.05$  & $0.875$   & $0.125$   & $0.1625$  & $0.20$  \\ \hline
f   & $-0.06$  & $0.0$     & $0.06$ \\
g   & $-0.05$  & $0.0$     & $0.05$ \\ 
h   & $-0.077$ & $-0.067$  & $-0.057$  & $-0.047$  & $-0.037$ & $-0.027$ & $-0.017$ & $0.003$ & $0.026$ & $0.063$ \\ 
i   & $-0.07$  & $-0.06$   & $-0.05$   & $-0.04$   & $-0.03$  & $-0.02$  & $-0.01$  & $0.0$   & $0.02$  & $0.04$  \\
\end{tabular}
\caption{\label{tab:partially_quenched_masses} Valence quark masses used in the configurations in \Tabref{tab:parameters_lattice_configs}. The second column denotes the case where sea quark mass and valence quark mass are equal, \ie full dynamic calculations, while the columns three to eleven mark partially quenched situations used for the pion pole extractions in \Secref{sec:technicalities_interpolation_pole_contributions_chiral_limit}.}
}


\subsection{Interpolation, pole contributions and chiral limit}\label{sec:technicalities_interpolation_pole_contributions_chiral_limit}

In the quenched case \cite{Gattringer:2004iv} we performed a chiral extrapolation of the renormalization factors using the different valance quark masses for each lattice setup and then stated these values as our final numbers for $Z_\Gamma$. In a full QCD calculation we only have one physical quark mass per lattice setup, so a chiral limit cannot be performed in a straightforward way. In this work we present the renormalization factors for each lattice setup and perform an extrapolation using different valence and sea quark masses, \ie a partially quenched approximation, only for discussing the continuum limit in \Secref{sec:results_collection} and for extracting the pion pole contributions.
The pseudoscalar density couples to the Goldstone boson channel, so we expect $\order{1/m}$ contributions to the pseudoscalar propagator in the chiral limit. As discussed in \cite{Gockeler:1998ye,Cudell:1998ic,Cudell:2001ny,Giusti:2000jr,Becirevic:2004ny} the inverse renormalization factor can be expanded as
\begin{equation}\label{eq:technicalities_z_p_functional_form}
  \frac{1}{Z_P} = \frac{A}{m} + B + C m + \order{m^2}
\end{equation}
and a ``subtracted'' renormalization factor can be defined by subtracting the pole term
\begin{equation}
  \frac{1}{Z_P^{Sub}} = \frac{1}{Z_P} - \frac{A}{m} \, .
\end{equation}
The operator product expansion ensures that pole contributions are suppressed for large values of $\mu$. For runs $a-e, \, h$ and $i$ we have sufficiently many different valence quark masses and we use the lowest five to perform a fit to \Eqref{eq:technicalities_z_p_functional_form}. For runs $f$ and $g$ we only have three valance quark masses, so here we only solve \Eqref{eq:technicalities_z_p_functional_form} for the coefficient. The valence quark masses are collected in \Tabref{tab:partially_quenched_masses}. \Figref{fig:Z_P_fit} shows the original and subtracted renormalization factors $Z_P$ and $Z_P^{Sub}$ of runs $h$ and $i$ for selected momentum vectors plotted against the AWI mass including the corresponding fit. \\
In \Figref{fig:Z_P_coeffs_vs_mu2} the coefficient $A$ of such a fit is displayed for runs $a-e,\, h$ and $i$. %
The coefficient $A$ is expected to be proportional to $\braket{\bar{q}\,q}/p^2$, where $-\braket{\bar{q}\,q} = \Sigma$ is the chiral condensate (cf.\, \cite{Noaki:2009xi,Aoki:2007xm}). In one-loop partially quenched chiral perturbation theory the condensate includes a logarithmic term that leads to a divergence in the chiral limit \cite{Golterman:1997st}, but a possible additional logarithmic term cannot be disentangled from the $1/m$ pole term with valance quark masses available.\footnote{Thanks to the referee for pointing this fact out.}%
Coefficient $A$ shows a rather universial behavior with a estimated uncertainty of $3\%$ for $\braket{\bar{q}\,q}$, which indicates a consistent pion-pole removal over the whole range of lattice configurations.
In order to compare runs at the same physical momentum transfer we use a cubic square interpolating fit. The values presented at momentum transfer $\mu=2\GeV$ are taken from such interpolating fits.\\
Uncertainties of all quantities are calculated using the standard jackknife procedure. Note, however that we include only uncertainties of statistical nature and no systematics due to finite size effects or other artifacts.\\
For selected momenta we have higher statistics available for run $a$ (23 configurations) and $h$ (15 configurations) from \Tabref{tab:parameters_lattice_configs}. As expected uncertainties are lower for larger ensembles, but in the region with $\mu > 1 \GeV$ all mean values for the renormalization factors with larger sample number lie within the error band of the results with smaller sample number. Only in the region below $1\GeV$ we find larger deviations, but those values also come with large uncertainties.
\FIGURE{
  \includegraphics[width=\plotwidth,clip]{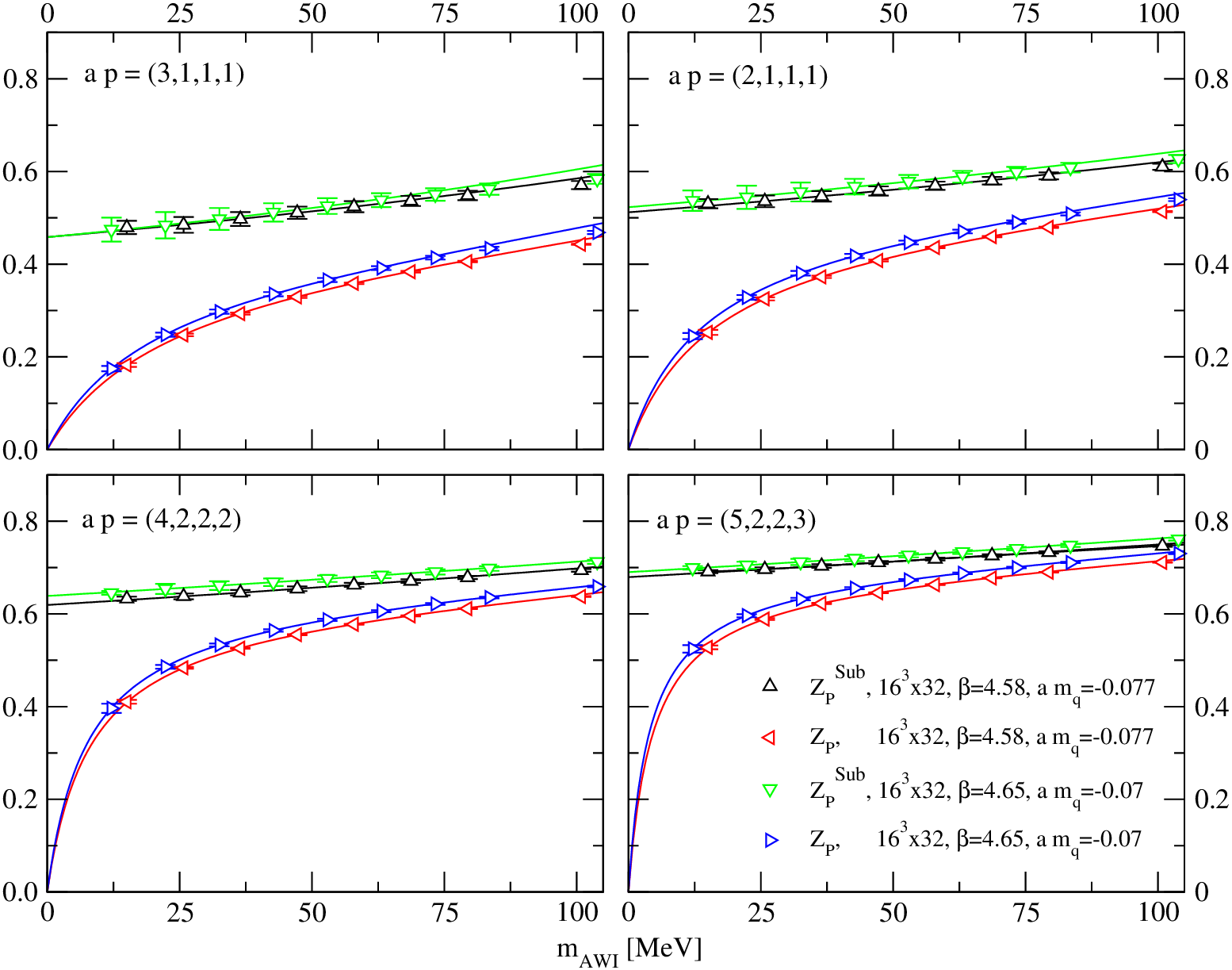}
  \caption{\label{fig:Z_P_fit}%
    The renormalization factors $Z_P$ and $Z_P^{Sub}$ plotted against the the AWI-mass for selected momenta of runs $h$ and $i$. %
    A fit to \Eqref{eq:technicalities_z_p_functional_form} is shown for $Z_P$ and the same fit with subtracted pole term is plotted for $Z_P^{Sub}$.
  }
}
\FIGURE{
  \includegraphics[width=\plotwidth,clip]{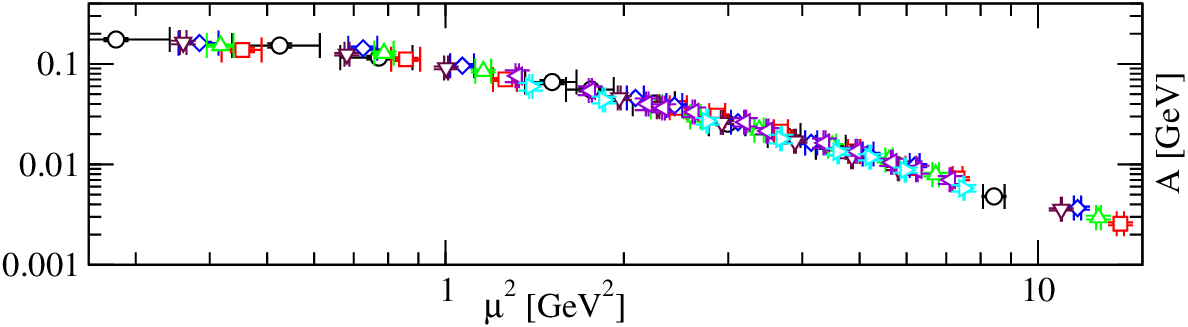}
  \caption{\label{fig:Z_P_coeffs_vs_mu2}The coefficient $A$ of the fit to \Eqref{eq:technicalities_z_p_functional_form} displayed against the $\mu^2$. We plot the runs %
  a (black circles), %
  b (red squares), %
  c (blue diamonds), %
  d (green upwards triangles), %
  e (brown downwards triangles) %
  on the $12^3\times24$ lattices and the runs %
  h (violet left pointing triangles) and %
  i (cyan right pointing triangles) %
  on the $16^3\times32$ lattices. %
  The parameters of the different runs are presented in \Tabref{tab:parameters_lattice_configs}.
}
}
%



\section{Results}\label{sec:results}

\subsection{Results in the \RIprime-scheme}\label{sec:results_ri_prime}

In \Figref{fig:results_Z_RI_prime_vs_mu2.12x24.regular} and \Figref{fig:results_Z_RI_prime_vs_mu2.16x32.regular} we present the result of $Z^{\RIprime}$ plotted against the momentum for runs on $12^3\times24$ and $16^3\times32$ lattices, respectively. Additionally an interpolating fit is displayed. Runs $b - e$ seem to agree for most of the quantities, whereas run $a$ lies closer to the runs on the $16^3\times32$ lattices. Therefore run $a$ is also included in the latter plot for comparison. This separation might be a hint for finite size effects, because run $a$ has a significantly larger box size then the rest of the  $12^3\times24$ setups and also the lattice spacing lies in the region of runs $f - i$.\\
A chiral operator is expected to have $Z_A = Z_V = \const$ and  $Z_P = Z_S = \const$. To this end we also plot the ratios $Z_P^{Sub,\RIprime} / Z_S^{\RIprime} = Z_P^{Sub} / Z_S$ and $Z_A^{\RIprime} / Z_V^{\RIprime}$ in \Figref{fig:results_Z_RI_prime_ratios_vs_mu2.16x32.regular}, where the scaling behavior is supposed to cancel out. In the interval $1.5\GeV < \mu < 3.0\GeV$ this is satisfied up to 7\% and 10\% for $Z_P^{Sub,\RIprime} / Z_S^{\RIprime}$ and $Z_A^{\RIprime} / Z_V^{\RIprime} = Z_A / Z_V$, respectively, on the larger lattices. The \DCI\, does not satisfying the GW-relation exactly, so we cannot expect the ratios to be equal to unity. In the quenched case \cite{Gattringer:2004iv} a comparable behavior was found for $Z_A / Z_V$, while $Z_P^{Sub} / Z_S$ was closer to unity. Note, that these ratios are the only quantities that we can directly compare to other studies, as the absolute values of the renormalization factors strongly depend on the formulation of the Dirac operator and the smearing method in use.
\FIGURE{
  \includegraphics[width=\plotwidthsmall,clip]{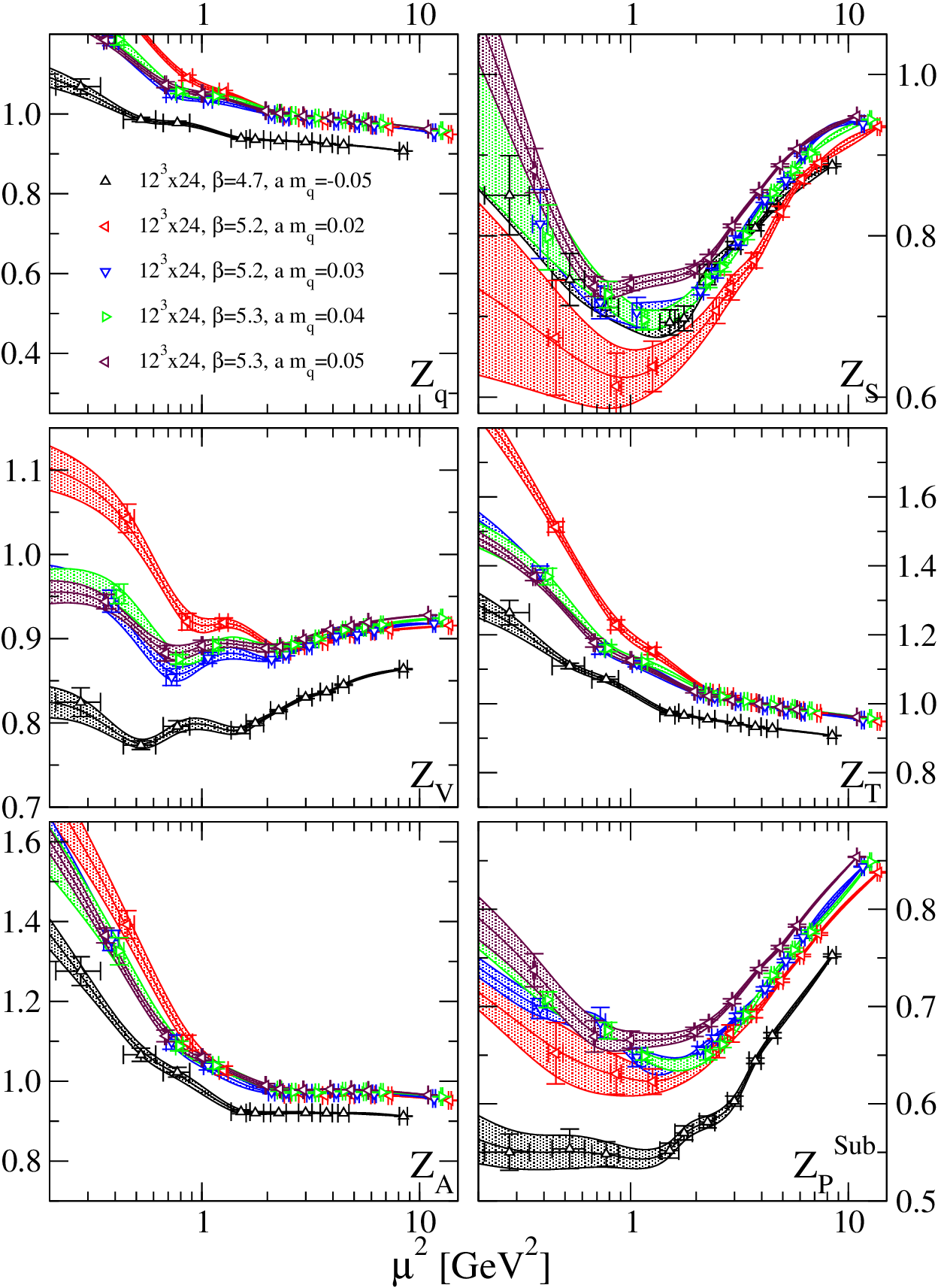}
  \caption{\label{fig:results_Z_RI_prime_vs_mu2.12x24.regular} The renormalization factor $Z^{\RIprime}$ displayed against $\mu^2$ for all $12^3\times24$ lattices.}
}
\FIGURE{
  \includegraphics[width=\plotwidthsmall,clip]{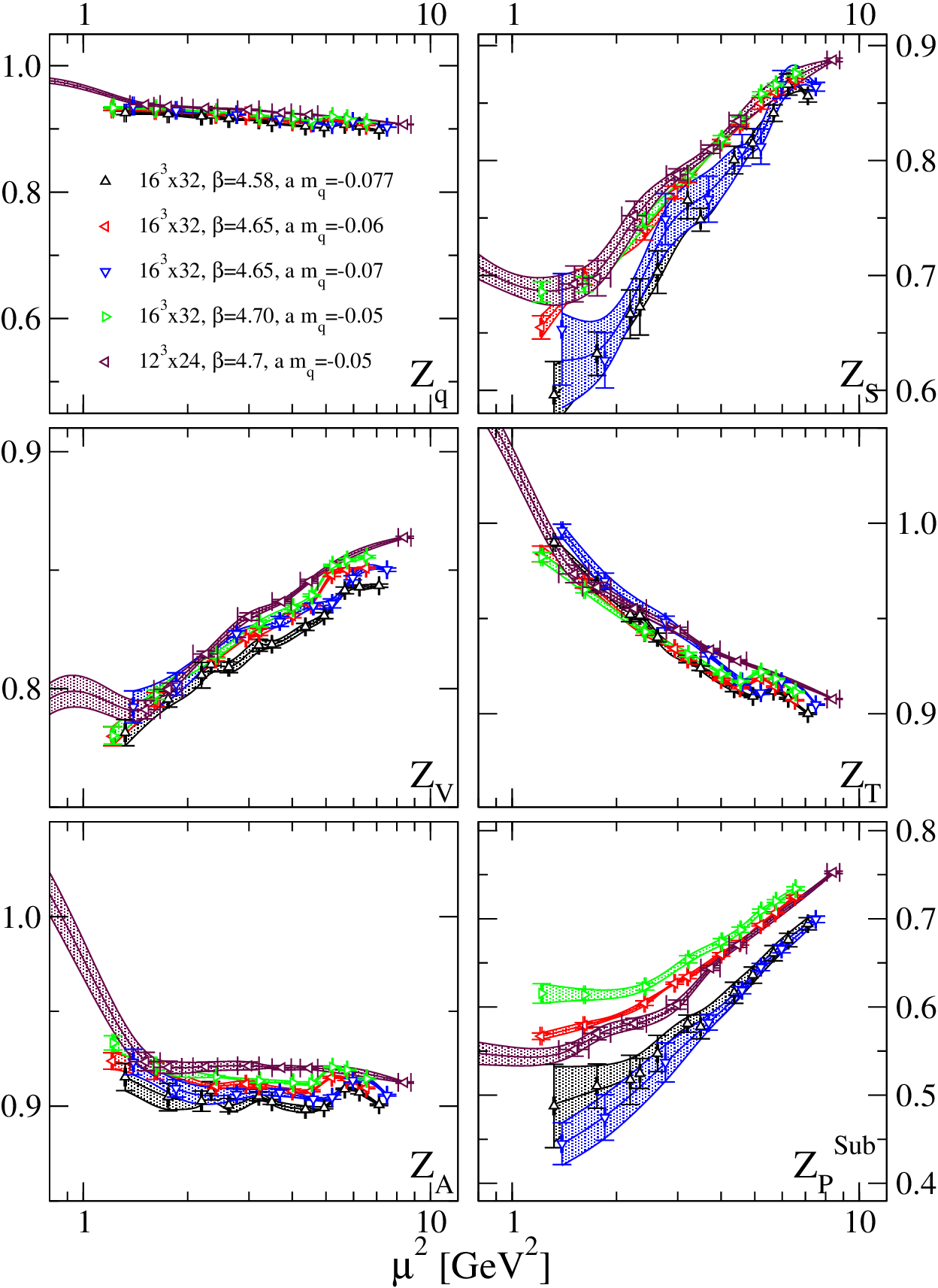}
  \caption{\label{fig:results_Z_RI_prime_vs_mu2.16x32.regular} The renormalization factor $Z^{\RIprime}$ displayed against $\mu^2$ for $12^3\times24$ with $\beta=4.7$ and all $16^3\times32$ lattice setups.}
}
\FIGURE{
  \includegraphics[width=\plotwidthsmall,clip]{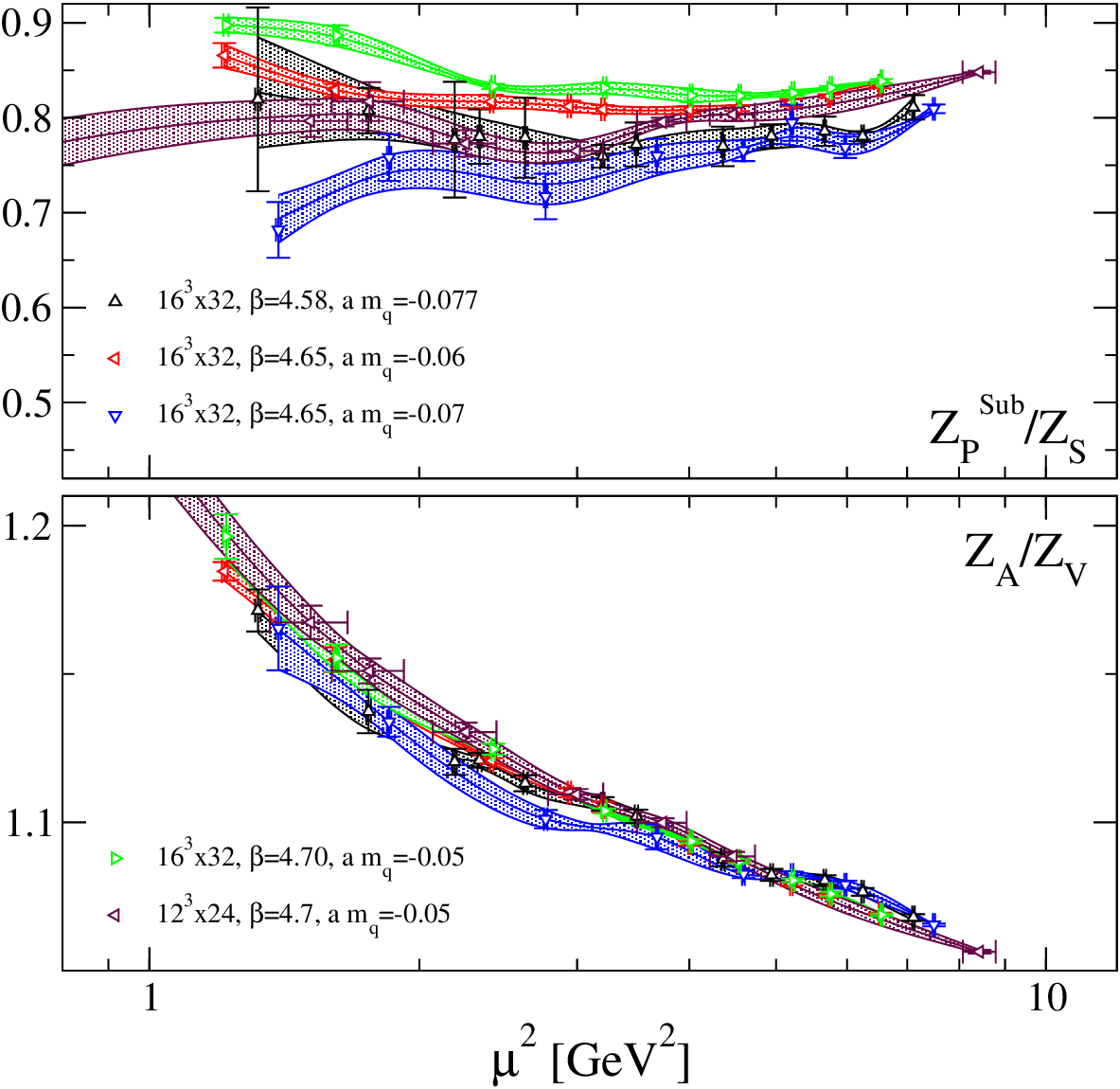}
  \caption{\label{fig:results_Z_RI_prime_ratios_vs_mu2.16x32.regular} The ratios $Z_P^{Sub} / Z_S$ and $Z_A / Z_V$ vs $\mu^2$ for $12^3\times24$ with $\beta=4.7$ and all $16^3\times32$ lattice setups.}
}
%


\subsection{Results in the \MSbar-scheme}\label{sec:results_ms}

The difference between the \RIprime-scheme we use and the \RI-scheme lies solely in the definition of the quark field renormalization factor, hence
\begin{equation}
  c_q^{\RIprime\RI} =\frac{Z_q^\RI}{Z_q^\RIprime} = \frac{Z_\Gamma^\RI}{Z_\Gamma^\RIprime} \, .
\end{equation}
The conversion factor expanded in terms of the strong coupling constant $\alpha_s$ in Landau gauge reads \cite{Franco:1998bm}
\begin{equation}\label{eq:conversion_c_ri_prime_ri}
\begin{split}
 c_q^{\RIprime\RI} =& 1 %
                     + \left\{ - \frac{67}{6} + \frac{2 N_f}{3} \right\} \left( \frac{\alpha_s}{4\pi} \right)^2 + \\
                    &+ \left\{ %
                         - \frac{52321}{72} + \frac{607 \zeta_3}{4} %
                         + \left( \frac{2236}{27} - 8 \zeta_3 \right) N_f %
                         - \frac{52}{27} N_f^2 %
                       \right\} \left( \frac{\alpha_s}{4\pi} \right)^3 %
                   + \order{\frac{\alpha_s}{4\pi}}^4 \, ,
\end{split}
\end{equation}
where $N_f$ denotes the number of dynamic quarks and $\zeta_n$ denotes the Riemann zeta function evaluated at $n$. \\
Due to Ward identities the conversion factor for axial vector and vector operators reads
\begin{equation}
  c_{V,A}^{\RIprime\MSbar} = c_\Gamma^{\RIprime\RI} c_{V,A}^{\RI\MSbar} = c_\Gamma^{\RIprime\RI} = c_q^{\RIprime\RI} \quad \textrm{\ie} \quad c_{V,A}^{\RI\MSbar} = 1 \, .
\end{equation}
For scalar and pseudoscalar sectors we have \cite{Chetyrkin:1999pq}
\begin{equation}
\begin{split}
  c_{P,S}^{\RI\MSbar} =& 1 + \frac{ 16 }{ 3 } \left( \frac{\alpha_s}{4\pi} \right) + \\
                       &+ \left\{ \frac{2246}{9} - \frac{89 N_f}{9} - \frac{152 \zeta_3}{3} \right\} \left( \frac{\alpha_s}{4\pi} \right)^2 + \\%
                       &+ \left\{ \frac{8290535}{648} - \frac{466375 \zeta_3}{108} + \frac{2960 \zeta_5}{9} \right. + \\
                       &     \hphantom{+} \hphantom{\{} + \left( -\frac{262282}{243} + \frac{4936 \zeta_3}{27} - \frac{80 \zeta_4}{3} \right) N_f + \\
                       &     \hphantom{+} \hphantom{\{} + \left. \left( \frac{8918}{729} + \frac{32 \zeta_3}{27} \right) N_f^2 \right\} %
                          \left( \frac{\alpha_s}{4\pi} \right)^3 %
                        + \order{\frac{\alpha_s}{4\pi}}^4\, .
\end{split}
\end{equation}
From \cite{Gracey:2000am,Gracey:2003yr} we get the conversion for the tensors
\begin{equation}
\begin{split}
  c_{T}^{\RI\MSbar} =& 1 %
                      + \left\{ -\frac{1622}{27} + \frac{184 \zeta_3}{9} + \frac{259 N_f}{81} \right\} \left( \frac{\alpha_s}{4\pi} \right)^2 + \\%
                     &+ \left\{ -\frac{15479317}{5832} + \frac{1209445 \zeta_3}{972} + \frac{1072 \zeta_4}{81} -\frac{10040 \zeta_5}{27} \right. + \\
                     &     \hphantom{+} \hphantom{\{} + \left( -\frac{1880 \zeta_3}{27} + \frac{80 \zeta_4}{9} + \frac{225890}{729} \right) N_f + \\
                     &     \hphantom{+} \hphantom{\{} + \left. \left( -\frac{32 \zeta_3}{81} - \frac{9542}{2187} \right) N_f^2 \right\} %
                        \left( \frac{\alpha_s}{4\pi} \right)^3 %
                      + \order{\frac{\alpha_s}{4\pi}}^4 \, .
\end{split}
\end{equation}
Finally for the quark field we use \cite{Gracey:2003yr}
\begin{equation}
\begin{split}
  c_{q}^{\RI\MSbar} =& 1 %
                      + \left\{  - \frac{1622}{27} + \frac{184 \zeta_3}{9} + \frac{259 N_f}{81} \right\} \left( \frac{\alpha_s}{4\pi} \right)^2 \\%
                     &+ \left\{ - \frac{15479317}{5832} + \frac{1209445 \zeta_3}{972} + \frac{1072 \zeta_4}{81} - \frac{10040 \zeta_5}{27} \right. \\
                     &     \hphantom{+} \hphantom{\{} + \left( \frac{225890}{729} - \frac {1880 \zeta_3}{27} + \frac {80 \zeta_4}{9} \right) N_f +  \\
                     &     \hphantom{+} \hphantom{\{} + \left. \left( -\frac {9542}{2187} - \frac{32 \zeta_3}{81} \right) N_f^2 \right\} %
                        \left( \frac{\alpha_s}{4\pi} \right)^3 %
                      + \order{\frac{\alpha_s}{4\pi}}^4 \, .
\end{split}
\end{equation}
The 3-loop expression for the coupling $\alpha_s$ in the \MSbar-scheme \cite{Alekseev:2002zn} reads
\begin{equation}
\begin{split}
  \frac{\alpha_s(q^2)}{4\pi} =& \frac{1}{\beta_0 \log(q^2)} %
                                - \frac{\beta_1}{\beta_0^3} \frac{\log(\log(q^2))}{\log(q^2)^2} \\%
                              &+ \frac{1}{\beta_0^5 \log(q^2)^3} \left( \beta_1^2 \log(\log(q^2))^2 - \beta_1^2 \log(\log(q^2)) + \beta_2 \beta_0 - \beta_1^2 \right) \, ,
\end{split}
\end{equation}
with $q=\mu/\Lambda_{\textrm{\scriptsize QCD}}$. The QCD scale (from \cite{Gockeler:2005rv} with $r_0 = 0.5 \fm$) in the \MSbar-scheme is given by
\begin{equation}
    \Lambda_{\textrm{\scriptsize QCD}}^{N_f=2} = 0.243(24)\GeV
\end{equation}
and we use the following coefficients \cite{vanRitbergen:1997va,Czakon:2004bu}
\begin{subequations} \label{eq:ms_beta_function_coefficients}
\begin{align}
  \beta_0             &= 11 - \frac{2}{3} N_f \, , \\
  \beta_1             &= 102 - \frac{38}{3}  N_f \, , \\
  \beta_2^{\MSbar}    &= \frac{2857}{2} - \frac{5033}{18} N_f + \frac{325}{54} N_f^2 \, , \\
  \begin{split}
    \beta_3^{\MSbar} =& \frac{149753}{6} + 3564 \zeta_3 + \left( -\frac{1078361}{162} - \frac{6508 \zeta_3}{27} \right) N_f \\
                      & + \left( -\frac{50065}{162} + \frac{6472 \zeta_3}{81} \right) N_f^2 + \frac{1093}{729} N_f^3 \, .
  \end{split}
\end{align}
\end{subequations}
The resulting renormalization factors in the \MSbar-scheme at $\mu=2\GeV$ are collected in \Tabref{tab:results_Z_MS_vs_mAWI_regular} and \Tabref{tab:results_Z_MS_vs_mAWI_reduced} for the original and the reduced definition, respectively.


\subsection{Results in the \RGI-form}\label{sec:results_rgi}

Renormalization factors in general depend on the renormalization scale $\mu$ in a way that is determined by the anomalous dimension $\gamma_O$ of the operator in consideration
\begin{equation} \label{eq:rgi_anomalous_dimension_definition}
  \gamma_O(\alpha_s) = -\mu \frac{d}{d\mu} \ln Z_O(\mu^2) = \sum_{i=0}^\infty \gamma_i \left( \frac{\alpha_s(\mu^2)}{4\pi} \right)^{i+1} \, .
\end{equation}
After integrating \Eqref{eq:rgi_anomalous_dimension_definition} we arrive at
\begin{equation}
  Z^\RGI = Z(\mu^2) \left( 2 \beta_0 \frac{\alpha_s(\mu^2)}{4\pi} \right)^{-\gamma_0/(2\beta_0)} \exp \left\{ \frac{1}{2} \int_0^{\alpha_s(\mu^2)} d\alpha \left( \frac{\gamma(\alpha)}{\beta(\alpha)} + \frac{\gamma_0}{\beta_0 \alpha} \right) \right\} \, ,
\end{equation}
the renormalization factor in \RGI-form (Renormalization Group Invariant), where the scale dependence coming from the renormalization group is extracted up to a certain order in perturbation theory. We use the the 4-loop expression of the QCD-$\beta$-function
\begin{equation}
  \beta = \frac{\mu}{2} \frac{d}{d\mu} \alpha_s(\mu^2) = -4 \pi \sum_i \beta_i \left( \frac{\alpha}{4\pi} \right)^{i+2} \, ,
\end{equation}
with coefficients listed in \Eqref{eq:ms_beta_function_coefficients}. Axial vector and vector are scale independent hence their anomalous dimensions are $\gamma_A = \gamma_V = 0$ and therefore $Z_{A,V}^{\RGI} = Z_{A,V}^{\MSbar}$. For $\gamma_q$, $\gamma_S = \gamma_P = \gamma_m^{-1}$ and $\gamma_T$ we use the 4-loop expressions from  \cite{vanRitbergen:1997va,Chetyrkin:1997dh,Gracey:2000am}.

In \Figref{fig:results_Z_RGI_vs_mu2.12x24.regular} and \Figref{fig:results_Z_RGI_vs_mu2.16x32.regular} the renormalization factors in \RGI-form are displayed against the momentum transfer $\mu^2$ for the $12^3\times24$ and $16^3\times32$ lattices, respectively. In \Figref{fig:results_Z_RGI_vs_mu2.16x32.reduced} the reduced renormalization factors in \RGI-form are presented against the same quantity. In the interval $1.5\GeV < \mu < 3.0\GeV$ we find a maximal deviation from the plateau behavior of 8\% for $Z_S$ and 5\% for the remaining renormalization factors. In the case of the reduced factors $\overline{Z}_q$, $\overline{Z}_V$, $\overline{Z}_A$ and $\overline{Z}_T$ we observe an almost linear scaling behavior in the region $\mu > 1.55\GeV$ or $a \mu > 1.15$. A linear fit of the form $c (1 + d (a \mu)^2)$ allows us to estimate discretization errors proportional to $(a \mu)^2$. For the afore mentioned quantities we find slope values $d$ of 0.019, 0.042, 0.020 and 0.039.
\FIGURE{
  \includegraphics[width=\plotwidth,clip]{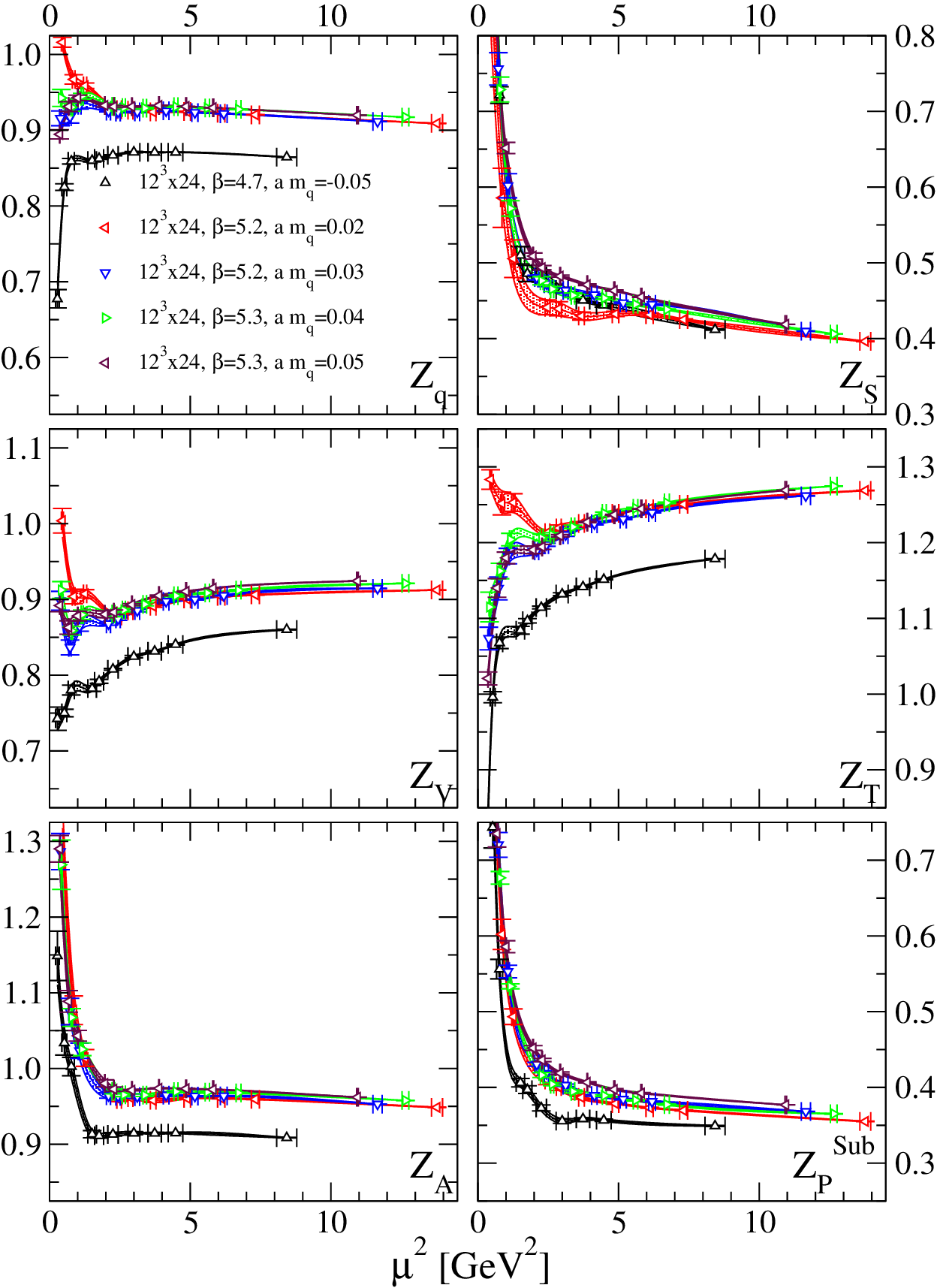}
  \caption{\label{fig:results_Z_RGI_vs_mu2.12x24.regular} The renormalization factors in \RGI-form $Z^{\RGI}$ plotted against $\mu^2$ for the $12^3\times24$ lattices, runs $a-e$ in \Tabref{tab:parameters_lattice_configs}.}
}
\FIGURE{
  \includegraphics[width=\plotwidth,clip]{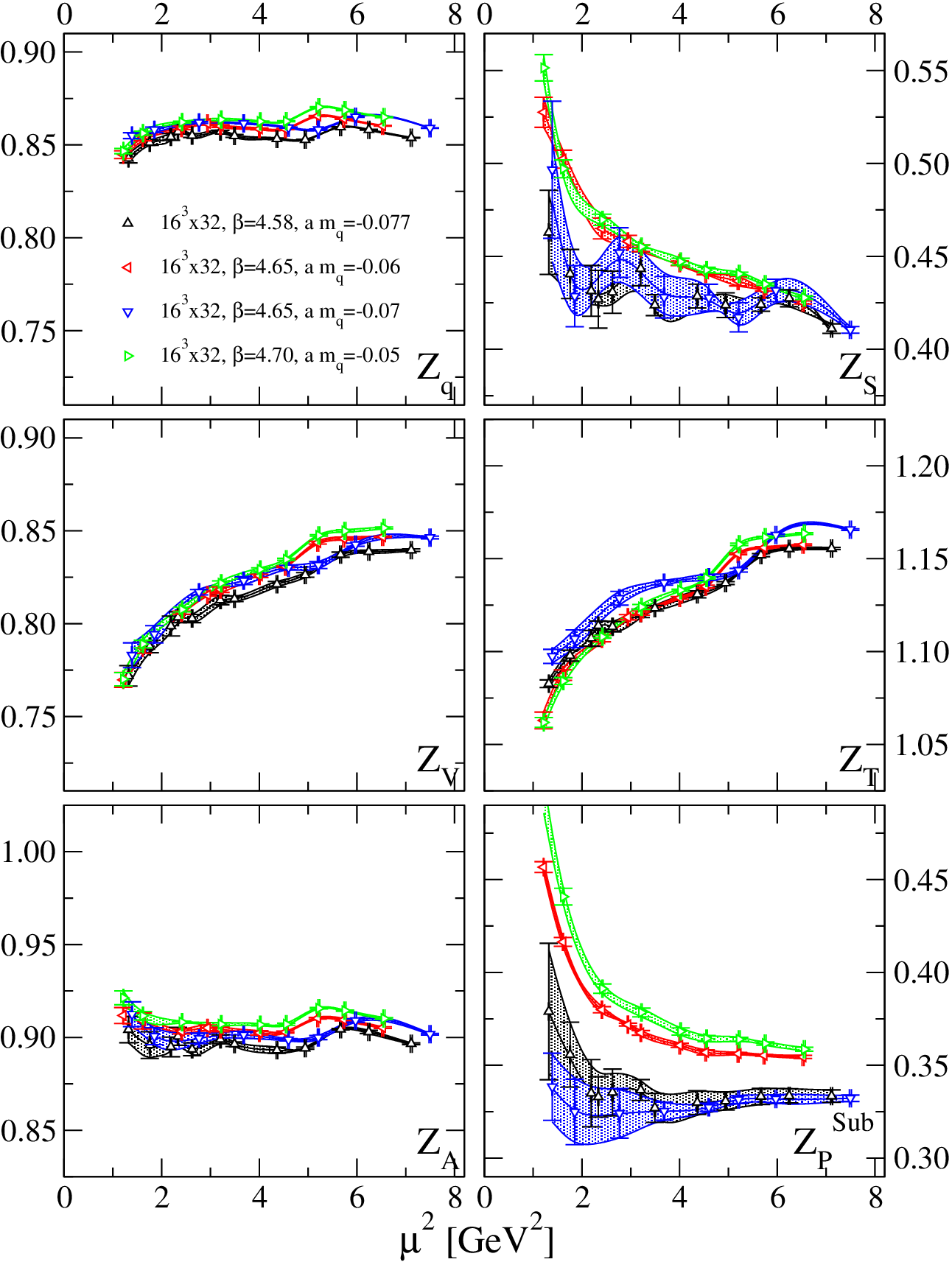}
  \caption{\label{fig:results_Z_RGI_vs_mu2.16x32.regular} The renormalization factors in \RGI-form $Z^{\RGI}$ plotted against $\mu^2$ for the $16^3\times32$ lattices, runs $f-i$ in \Tabref{tab:parameters_lattice_configs}.}
}
\FIGURE{
  \includegraphics[width=\plotwidth,clip]{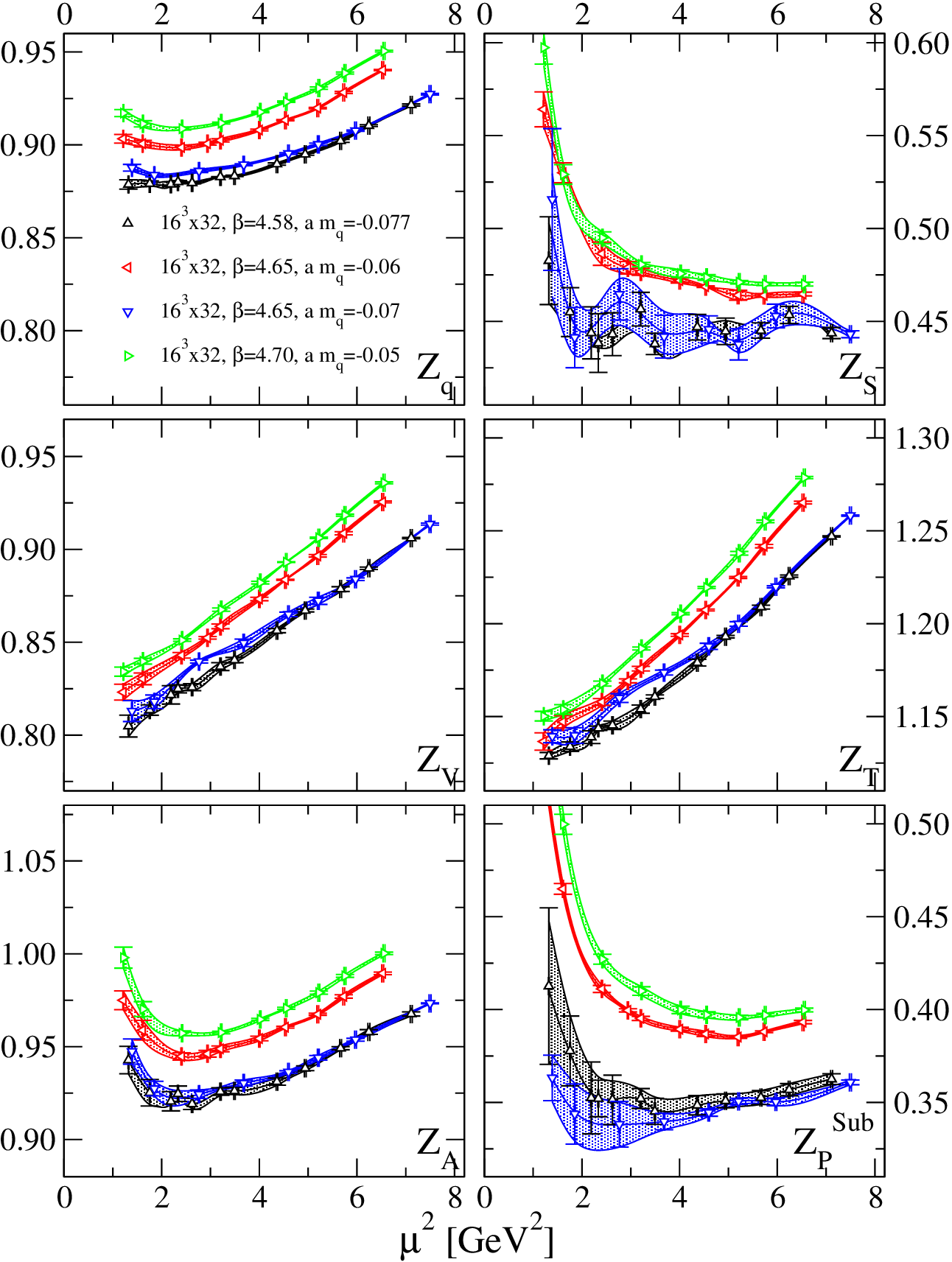}
  \caption{\label{fig:results_Z_RGI_vs_mu2.16x32.reduced} The reduced renormalization factors in \RGI-form $\overline{Z}^{\RGI}$ plotted against $\mu^2$ for the $16^3\times32$ lattices, runs $f-i$ in \Tabref{tab:parameters_lattice_configs}.}
}


\subsection{Collection of Results}\label{sec:results_collection}

In this section we collect results of renormalization factors for the original and the reduced method as well as for massive quarks and a chiral extrapolation in \Tabsref{tab:results_Z_MS_vs_mAWI_regular}{tab:results_Z_MS_vs_mAWI_reduced_chiral}. The same quantities are plotted in \Figref{fig:results_Z_MS_vs_a} against the lattice spacing. The difference between the original and the reduced method lies in the subtraction of the component of the quark propagator that is proportional to the unit matrix for the latter method (cf.~\Eqref{eq:quark_propagator_definition_reduced}). Thereby a potential cut-off artefact is reduced according to \cite{Capitani:2000xi}. %
It is interesting to note that the renormalization factors scalar, vector, tensor and axial vector in the reduced definition differ from the original definition solely by a factor $\overline{Z}_q/Z_q$ before taking the chiral limit, which means that the denominator in \Eqref{eq:Z_Gamma_definition_reduced} is not influenced by the redefinition of the quark propagator. For the pseudoscalar renormalization factor we observe a maximum deviation of 5\% from this behavior. For chirally extrapolated values this strict factorization in no longer observed, but still holds approximately, with deviations in the sub percent range for vector, tensor and axial vector, and below 3 \% and 5 \% for the scalar and pseudoscalar, respectively.

\TABLE{
\begin{tabular}{c|c||l|l|l|l|l|l}
  \mAWI [\MeV] & $a[\textrm{fm}]$ & $Z_q$     &  $Z_S$    &  $Z_V$    &  $Z_T$     &  $Z_A$     &  $Z_P^{Sub}$ \\ \hline
%
  31(4)         & 0.147(18)      &  0.9116(3)    &  0.973(4)     &  0.8345(8)    &  0.9094(7)    &  0.915(1)     &  0.779(3) \\
  43(2)         & 0.115(6)       &  0.9672(8)    &  0.93(1)      &  0.893(1)     &  0.976(1)     &  0.958(2)     &  0.833(4) \\
  58(3)         & 0.125(6)       &  0.9689(7)    &  0.997(4)     &  0.896(1)     &  0.972(1)     &  0.965(2)     &  0.849(2) \\
  61(2)         & 0.120(4)       &  0.9739(9)    &  0.989(5)     &  0.900(2)     &  0.978(1)     &  0.968(1)     &  0.850(3) \\
  76.5(6)       & 0.129(1)       &  0.9763(3)    &  1.021(2)     &  0.9060(6)    &  0.9758(5)    &  0.9739(4)    &  0.881(2) \\
\hline
%
  35.0(3)              & 0.150(1)            &  0.8987(5)    &  0.969(3)     &  0.8257(9)    &  0.8967(5)    &  0.902(1)     &  0.783(2) \\
  43.1(5)              & 0.150(2)            &  0.9030(3)    &  0.972(4)     &  0.8290(9)    &  0.8998(4)    &  0.9068(5)    &  0.800(4) \\
  15.0(4)              & 0.144(2)            &  0.8934(7)    &  0.93(1)      &  0.818(2)     &  0.896(2)     &  0.894(2)     &  0.71(1) \\
  12.1(5)              & 0.140(1)            &  0.9013(3)    &  0.93(2)      &  0.826(2)     &  0.9042(8)    &  0.901(1)     &  0.707(8) \\
\end{tabular}
\caption{\label{tab:results_Z_MS_vs_mAWI_regular}The interpolated renormalization factors in the \MSbar-scheme $Z^{\MSbar}$ with the corresponding AWI mass $\mAWI$ at momentum transfer $\mu = 2\GeV$.}
}
\TABLE{
\begin{tabular}{c|c||l|l|l|l|l|l}
  \mAWI [\MeV] & $a[\textrm{fm}]$ & $Z_q$     &  $Z_S$    &  $Z_V$    &  $Z_T$     &  $Z_A$     &  $Z_P^{Sub}$ \\ \hline
%
  31(4)         & 0.147(18)      &  0.961(1)     &  1.025(5)     &  0.8798(9)    &  0.959(1)     &  0.964(2)     &  0.855(3) \\
  43(2)         & 0.115(6)       &  1.005(2)     &  0.97(1)      &  0.928(2)     &  1.014(2)     &  0.995(3)     &  0.912(5) \\
  58(3)         & 0.125(6)       &  1.027(2)     &  1.057(4)     &  0.950(2)     &  1.030(2)     &  1.023(2)     &  0.930(2) \\
  61(2)         & 0.120(4)       &  1.031(1)     &  1.047(5)     &  0.953(2)     &  1.035(2)     &  1.024(2)     &  0.929(3) \\
  76.5(6)       & 0.129(1)       &  1.0477(7)    &  1.096(2)     &  0.9722(9)    &  1.047(1)     &  1.0452(9)    &  0.976(2) \\
\hline
%
  35.0(3)               & 0.150(1)              &  0.9504(5)    &  1.025(3)     &  0.8732(9)    &  0.9483(6)    &  0.954(1)     &  0.846(2) \\
  43.1(5)               & 0.150(2)              &  0.9604(4)    &  1.033(4)     &  0.8817(7)    &  0.9570(4)    &  0.9645(6)    &  0.869(4) \\
  15.0(4)               & 0.144(2)              &  0.9278(6)    &  0.96(1)      &  0.849(1)     &  0.930(1)     &  0.928(2)     &  0.75(1) \\
  12.1(5)               & 0.140(1)              &  0.9327(4)    &  0.96(2)      &  0.855(2)     &  0.9358(9)    &  0.932(1)     &  0.740(7) \\
\end{tabular}
\caption{\label{tab:results_Z_MS_vs_mAWI_reduced}The reduced, interpolated renormalization factors in the \MSbar-scheme $\overline{Z}^{\MSbar}$ with the corresponding AWI mass $\mAWI$ at momentum transfer $\mu = 2\GeV$.}
}
\TABLE{
\begin{tabular}{c|c||l|l|l|l|l|l}
  \mAWI [\MeV] & $a[\textrm{fm}]$ & $Z_q$     &  $Z_S$    &  $Z_V$    &  $Z_T$     &  $Z_A$     &  $Z_P^{Sub}$ \\ \hline
%
  31(4)         & 0.147(18)      &  0.9007(3)    &  0.948(4)     &  0.8220(7)    &  0.8992(7)    &  0.903(1)     &  0.748(4) \\
  43(2)         & 0.115(6)       &  0.9559(8)    &  0.921(5)     &  0.880(1)     &  0.965(1)     &  0.943(2)     &  0.806(4) \\
  58(3)         & 0.125(6)       &  0.9519(8)    &  0.958(3)     &  0.877(1)     &  0.957(1)     &  0.945(2)     &  0.817(3) \\
  61(2)         & 0.120(4)       &  0.957(1)     &  0.949(2)     &  0.882(2)     &  0.964(2)     &  0.949(1)     &  0.820(4) \\
  76.5(6)       & 0.129(1)       &  0.9532(4)    &  0.976(1)     &  0.8814(6)    &  0.9563(6)    &  0.9489(5)    &  0.841(2) \\
\hline
%
  35.0(3)                 & 0.150(1)              &  0.8860(5)    &  0.943(2)     &  0.8112(7)    &  0.8856(6)    &  0.8883(7)    &  0.761(3) \\
  43.1(5)                 & 0.150(2)              &  0.8880(2)    &  0.945(1)     &  0.812(1)     &  0.8873(4)    &  0.8910(6)    &  0.774(5) \\
  15.0(4)                 & 0.144(2)              &  0.8887(7)    &  0.918(7)     &  0.813(2)     &  0.892(2)     &  0.889(1)     &  0.69(2) \\
  12.1(5)                 & 0.140(1)              &  0.8974(3)    &  0.920(6)     &  0.822(2)     &  0.9015(7)    &  0.896(1)     &  0.70(2) \\
\end{tabular}
\caption{\label{tab:results_Z_MS_vs_mAWI_regular_chiral}A chiral extrapolation for the interpolated renormalization factors in the \MSbar-scheme $Z^{\MSbar,\textrm{chir}}$ with the corresponding AWI mass $\mAWI$ at momentum transfer $\mu = 2\GeV$.}
}
\TABLE{
\begin{tabular}{c|c||l|l|l|l|l|l}
  \mAWI [\MeV] & $a[\textrm{fm}]$ & $Z_q$     &  $Z_S$    &  $Z_V$    &  $Z_T$     &  $Z_A$     &  $Z_P^{Sub}$ \\ \hline
%
  31(4)         & 0.147(18)      &  0.925(1)     &  0.969(5)     &  0.8453(9)    &  0.925(1)     &  0.928(2)     &  0.799(4) \\
  43(2)         & 0.115(6)       &  0.957(1)     &  0.891(6)     &  0.878(2)     &  0.967(2)     &  0.940(3)     &  0.849(4) \\
  58(3)         & 0.125(6)       &  0.962(1)     &  0.956(4)     &  0.884(2)     &  0.967(2)     &  0.954(2)     &  0.868(3) \\
  61(2)         & 0.120(4)       &  0.965(1)     &  0.939(3)     &  0.886(2)     &  0.972(2)     &  0.955(2)     &  0.867(4) \\
  76.5(6)       & 0.129(1)       &  0.9636(6)    &  0.971(2)     &  0.8903(8)    &  0.9672(9)    &  0.9590(7)    &  0.896(2) \\
\hline
%
  35.0(3)                 & 0.150(1)              &  0.9091(5)    &  0.959(2)     &  0.8267(7)    &  0.9087(6)    &  0.9061(8)    &  0.806(2) \\
  43.1(5)                 & 0.150(2)              &  0.9115(3)    &  0.960(1)     &  0.8314(8)    &  0.9107(4)    &  0.9134(7)    &  0.819(5) \\
  15.0(4)                 & 0.144(2)              &  0.9124(6)    &  0.941(7)     &  0.834(2)     &  0.916(1)     &  0.912(2)     &  0.72(2) \\
  12.1(5)                 & 0.140(1)              &  0.9202(4)    &  0.942(6)     &  0.842(1)     &  0.9237(7)    &  0.919(1)     &  0.72(2) \\
\end{tabular}
\caption{\label{tab:results_Z_MS_vs_mAWI_reduced_chiral}A chiral extrapolation for the reduced, interpolated renormalization factors in the \MSbar-scheme $\overline{Z}^{\MSbar,\textrm{chir}}$ with the corresponding AWI mass $\mAWI$ at momentum transfer $\mu = 2\GeV$.}
}

\FIGURE{
  \includegraphics[width=\plotwidth,clip]{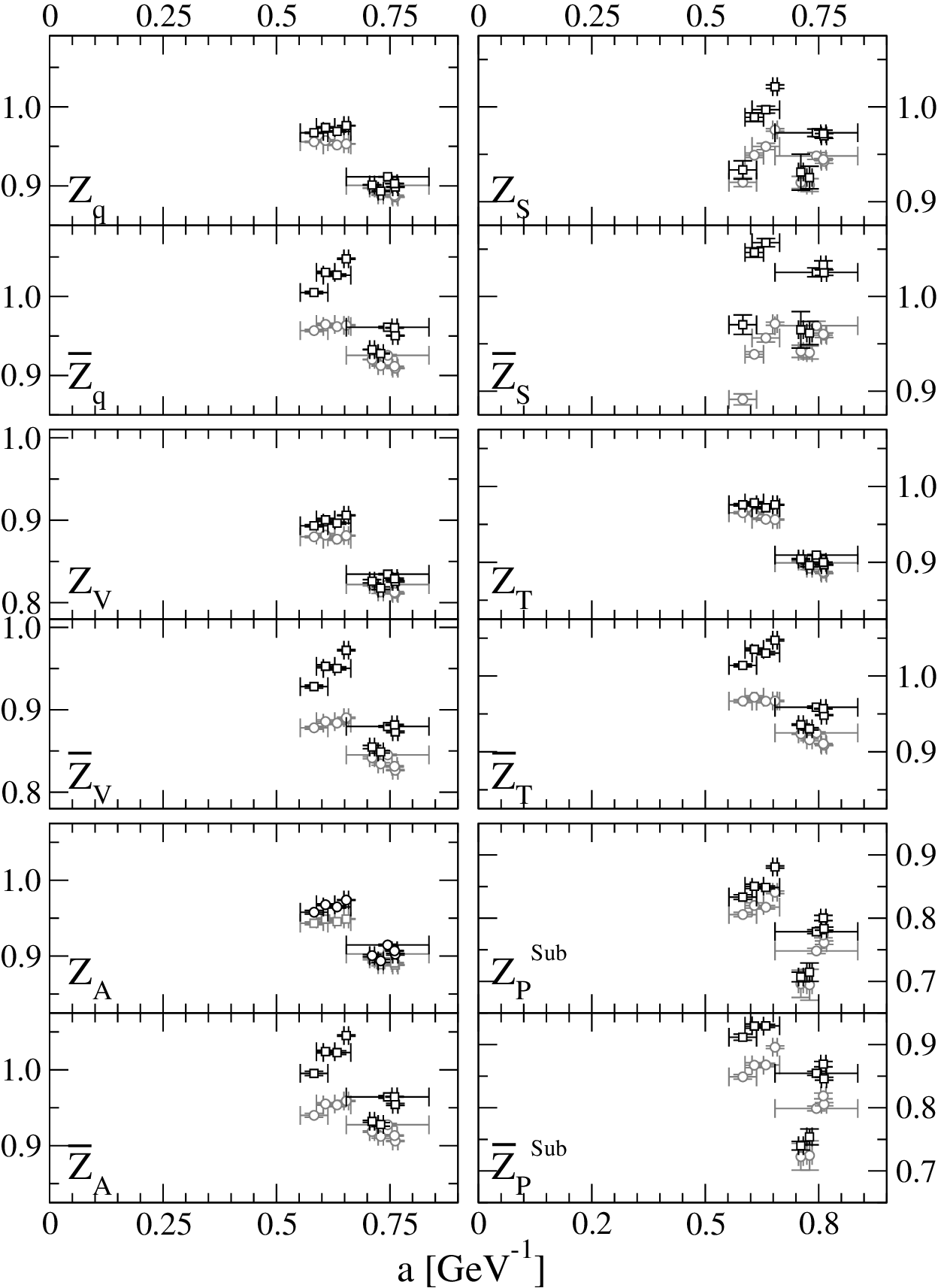}
  \caption{\label{fig:results_Z_MS_vs_a} The interpolated renormalization factors in the \MSbar-scheme $Z^{\MSbar}$ and $\overline{Z}^{\MSbar}$ at momentum transfer $\mu = 2\GeV$ plotted against the lattice spacing $a$ for massive quarks (black) and a chiral extrapolation (gray).}
}


\section{Summary and Conclusion}\label{sec:summary_and_conclusion}

Renormalization factors are essential to relate computations of renormalization scheme dependent quantities like the pion decay constant and quark mass on the lattice with results from continuum calculations or measurements. We present values of the renormalization factors for quark bilinears in the regularization independent scheme and the modified minimal subtraction scheme using conventions of \cite{Martinelli:1995ty,Gockeler:1998ye} and \cite{Capitani:2000xi}. For the pseudoscalar factor the pion pole is subtracted and a chiral extrapolation was performed.


\clearpage


\acknowledgments

This work is based on gauge configurations obtained from the BGR collaboration on the SGI Altix 4700 of the Leibniz-Rechenzentrum Munich. %
The quark propagators have been determined on the Sun Fire V20z cluster of the computer center of Karl-Franzens-Universit\"at Graz. \\%
I am deeply grateful for the support Christian Lang gave me and for our many interesting discussions. I also appreciate the help of Meinulf G\"ockeler very much.


\appendix





\providecommand{\href}[2]{#2}\begingroup\raggedright\endgroup



\end{document}